\documentclass[conference]{IEEEtran}
\IEEEoverridecommandlockouts
\usepackage{cite}
\usepackage{amsmath,amssymb,amsfonts}
\usepackage{algorithmic}
\usepackage{graphicx}
\usepackage{textcomp}
\usepackage{xcolor}

\usepackage{subfigure}
\usepackage[english]{babel}
\usepackage[utf8x]{inputenc}
\usepackage{caption}
\usepackage[normalem]{ulem}
\usepackage{array}
\usepackage{multirow}
\usepackage{booktabs}

\newcommand{\mxj}[1]{{\color{black} #1}}
\newcommand{\xm}[1]{{\color{black} #1}}

\newcommand{\zhenyi}[1]{{\color{black} #1}}
\newcommand{\FY}[1]{{\color{black} #1}}

\newcommand{\gabe}[1]{{\color{black} #1}}

\newcommand{\Karl}[1]{{\color{black} #1}}

\newcolumntype{R}[1]{>{\raggedleft\arraybackslash }m{#1}}
\newcolumntype{L}[1]{>{\raggedright\arraybackslash }m{#1}}
\newcolumntype{C}[1]{>{\centering\arraybackslash }m{#1}}

\def\BibTeX{{\rm B\kern-.05em{\sc i\kern-.025em b}\kern-.08em
    T\kern-.1667em\lower.7ex\hbox{E}\kern-.125emX}}
\begin{document}

\title{Manifest the Invisible: Design for Situational Awareness of Physical Environments in Virtual Reality
}

\author{
 \IEEEauthorblockN{1\textsuperscript{st} Zhenyi He}
 \IEEEauthorblockA{\textit{New York University} \\
 zh719@nyu.edu}
 \and
 \IEEEauthorblockN{2\textsuperscript{nd} Fengyuan Zhu}
 \IEEEauthorblockA{\textit{New York University} \\
 zhufyaxel@gmail.com }
 \and
 \IEEEauthorblockN{3\textsuperscript{rd} Ken Perlin}
 \IEEEauthorblockA{\textit{New York University} \\
 ken.perlin@gmail.com}
 \and
 \IEEEauthorblockN{4\textsuperscript{th} Xiaojuan Ma}
 \IEEEauthorblockA{\textit{The Hong Kong University of Science and Technology} \\
 xiaojuan.c.ma@gmail.com }
}

\maketitle

\begin{abstract}
Virtual Reality (VR) provides immersive experiences in the virtual world, but it may reduce users’ awareness of physical surroundings and cause safety concerns and psychological discomfort. Hence, there is a need of an ambient information design to increase users’ situational awareness (SA) of physical elements when they are immersed in VR environment. This is challenging, since there is a tradoff between the awareness in reality and the interference with users' experience in virtuality. 
In this paper, we design five representations (indexical, symbolic, and iconic with three emotions) based on two dimensions (vividness and emotion) to address the problem. We conduct an empirical study to evaluate participants’ SA, perceived breaks in presence (BIPs), and perceived engagement through VR tasks that require movement in space. Results show that designs with higher vividness evoke more SA, designs that are more consistent with the virtual environment can mitigate the BIP issue, and emotion-evoking designs are more engaging.
\end{abstract}


\section{Introduction}
Creating the sense of immersion and presence in the virtual world is critical in Virtual Reality (VR) design\cite{heeter1992being}. Due to the fact that virtual and physical realities overlap with each other in conceptualization and implementation~\cite{bowman2007virtual}, things occurring to users in the real environment can affect their experiences in the virtual environment (VE). Given that VR users' situational awareness(SA) of the real world diminishes as a result of obscured vision, confusion between reality and virtuality, and dynamics of the real environment\cite{endsley2016designing}, these may confront various types of discomfort in the VR interaction. 
First, users are isolated in an environment whose representation may not align with their prior knowledge of the space~\cite{Benford:2013:UUE:2500468.2500889}.
If users come into contact with elements in the physical environment that are not rendered in or mapped to the VR world by design, they might get confused or even injured. A recent news article showed that a man accidentally killed himself by running into furniture while playing a VR game, as he lacked feedback \FY{to keep himself safe}~\cite{Kim_2017}. 
Bearing such risks in mind, users may feel disempowered as they surrender control to the VR system\cite{Benford:2013:UUE:2500468.2500889}. 
Second, if users are playing with the VR device in a public venue, they are likely to be watched by others not present in the virtual scene, introducing ``a sense of vulnerability inherent in surveillance by unseen observers'' \cite{Benford:2013:UUE:2500468.2500889}. Moreover, VR users have little knowledge of and control over the way an audience impacts their physical presence.
They might be subjected to the audience’s capricious or mischievous behaviors, such as tricking and taking likely embarrassing photos. 

This actually presents a perplexing dilemma in VR interaction design. On the one hand, if we do not improve VR users' SA of the physical world, the entailing discomfort as mentioned above is likely to disturb users' immersion in virtuality, and even engenders breaks in presence (BIPs) \cite{garau2008temporal,slater2000virtual}.
On the other hand, if users are constantly made aware of their physical experience, then they may lose focus on the alternative virtual reality. As a result, they will experience BIP.This raises an important question of how to ensure sufficient SA to minimize potential discomfort in VR interactions without breaking the virtual experiences.

One possible solution is to project the consciousness of the external world onto the representation of the virtual scene~\cite{sra2016asymmetric}; in other words, transforming situational awareness to mediate perception of the environment (a.k.a. environmental presence \cite{heeter1992being,schuemie2001research}) as well as other beings in it (a.k.a. social presence \cite{heeter1992being,schuemie2001research}). Conventionally, social presence and environmental presence in VR refer to the salience of and interaction with characters and objects included in the VE~\cite{sra2016asymmetric}. However, little work has been done to answer the question: if irrelevant ``reality'' breaks through the boundary of the two worlds, how to raise users' awareness without transporting their consciousness back to the physical surrounding, i.e., avoiding BIP?

In this paper, we apply the research through design (RtD) method~\cite{Zimmerman2007RTD} and propose to promote users' SA of physical object(s) and human being(s) in reality when they break into the VE, by rendering them as part of the environmental and social presence in the digital world.Through a within-subject study with 25 participants, we compared the efficacy of five design\Karl{s} in five representational fidelities, i.e., indexical, symbolic, iconic with a positive emotion, iconic with a neutral emotion, and iconic with a negative emotion, on raising VR users' SA  of the physical world while maintaining consciousness in the VE. To the best of our knowledge, this is the first technical discussion on the design of virtual presence of physical elements in VR.
Our main contributions are as follows.
\begin{itemize}
\item We explore the design of situational awareness to physical environmental factors and design five representational fidelities of presence based on aesthetic theories.
\item We take a research through design (RtD) approach and experiment with two properties of representation (vividness and emotion).
\item We evaluate proposed hypotheses through a within-subject user study, and provide guidelines for designing virtual presence of environmental elements in virtual reality.
\end{itemize}

 \section{Related Work}


\subsection{Situational Awareness in VR}
First, we identify situational awareness and look into the approach to present and measure situational awareness.
Situational awareness is the ability to identify, process, and comprehend critical elements of information of what is happening with regards to a goal, or in other words, what is going on in the environment.
It can be measured in three levels\cite{endsley2016designing}.
Level 1: Perception of elements in the environment.
Level 2: Comprehension of the current situation.
and Level 3: Projection of future status.
Situational awareness is the key to user-centered design.
In terms of immersive experiences like VR, discomfort will be introduced by low situational awareness in VR.



\subsection{Presence in VR}
``Presence is consciousness in that virtual reality''\cite{sanchez2005presence}. Such saying refers to three types of presence,
(1) Personal presence, (2) Social presence, and (3) Environmental presence. 
For personal presence, it could be measured in various approaches~\cite{heeter1992being}, and breaks in presence (BIP) is one important issue that might happen during such experience.
BIP is defined as any perceived phenomenon during the exposure to VR that instigates users' awareness of real-world setting during the experience, `breaking' their personal presence in the VE \cite{sanchez2005presence,schuemie2001research}. 
Social presence means ``the degree of salience of others in a mediated communication''~\cite{short1976social}. Like what ~\cite{torisu_2016} presented, social presence can be enhanced by implementing multi-play (users can see each other) or different timelines (showing ``being there with others'').
Environmental presence does not refer to one's surroundings as they exist in the physical world, but to the perception of those surroundings as mediated by both automatic and controlled mental processes~\cite{steuer1992defining}.
In addition, \zhenyi{ there are two dimensions which affect a lot in communication technologies during telepresence: vividness and interactivity~\cite{steuer1992defining}.}
Vividness is the ability to induce a sense of presence. It means the representational richness of a mediated environment as defined by its formal features~\cite{steuer1992defining}. 
Interactivity is how users can participate in revising the form and content~\cite{steuer1992defining}.

\subsection{Presence of physical objects in VR}



Prior research has shown that, by careful calibration, VR designers can deliberately incorporate certain activities in the physical world, such as real walking~\cite{iwata2005circulafloor,sun2016mapping}, drawing~\cite{otsuki2010mai,sugihara2011mai} and actually touching the physical entity of a virtual object \cite{stone2001haptic} as a part of social and environmental presence, to enhance the sense of immersion and presence in virtuality. 
\zhenyi{Sun et al. discussed the mapping between physical world and virtual world~\cite{sun2016mapping}. Considering different room sizes, wall shapes, and surrounding objects in the virtual and real worlds, it attempted to warp the virtual world appearance into real world geometry, for example how a physical table became a virtual wall in users' VR experience.
}

\zhenyi{However, prior work mostly focused on (1) presenting the physical objects including passive objects~\cite{liao2000force} and actuated systems like NormalTouch~\cite{benko2016normaltouch}, PhyShare~\cite{He2017PhyShareSP}, SnakeCharmer~\cite{araujo2016snake} and TurkDeck~\cite{cheng2017mutual, cheng2015turkdeck, cheng2014haptic} in VR for haptic feedback or direct manipulation;
Or (2) presenting human being as another player in VR for social interaction and collaboration research\cite{hoyer2004multiuser}, which has different identity from our work.
NormalTouch~\cite{benko2016normaltouch} provided direct manipulation through the physical objects.
PhyShare~\cite{He2017PhyShareSP} created different mapping between virtual proxy and physical robots and controlled the robots to provide instant haptic feedback to indicate the existence of physical objects. It visualized the object by similar representation.
SnakeCharmer~\cite{araujo2016snake} offered different texture to mimic different objects so that users felt differently when touching it. All objects are rendered as cubes which exactly the same as physical object itself.
TurkDeck~\cite{cheng2017mutual, cheng2015turkdeck, cheng2014haptic} is a multiuser experience, however users play as main-actor in their own VR experience and play as part of the environment in other's scenarios. The design of the experiment avoids users to have interactions with each other.
}


\section{Design for Environmental Factors}
 \subsection{Design Considerations}

\begin{figure*}[h]
  \centering
  \subfigure[indexical design]{\includegraphics[width=0.19\textwidth]{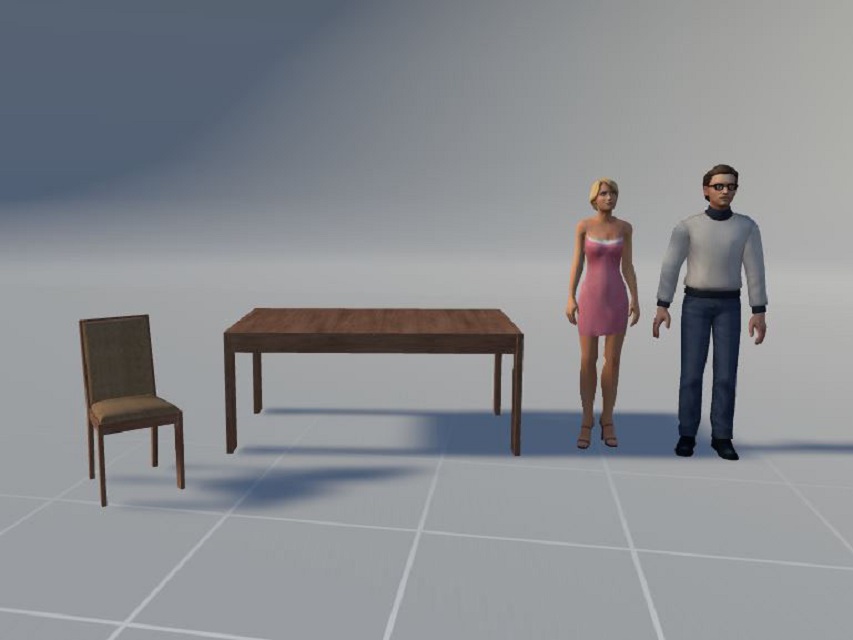}\label{fig:designs:a}}
  \subfigure[symbolic design]{\includegraphics[width=0.19\textwidth]{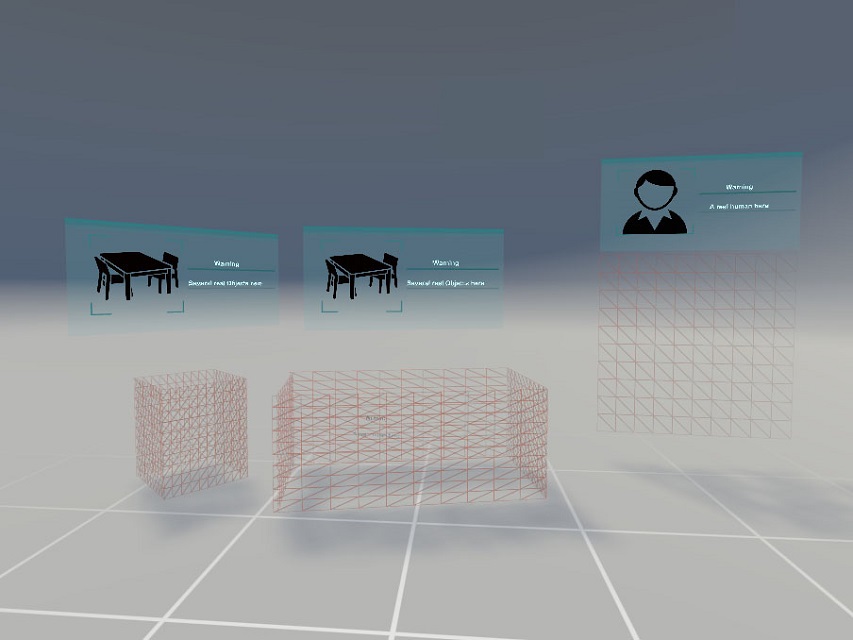}\label{fig:designs:b}}
  \subfigure[iconic design with positive emotion]{\includegraphics[width=0.19\textwidth]{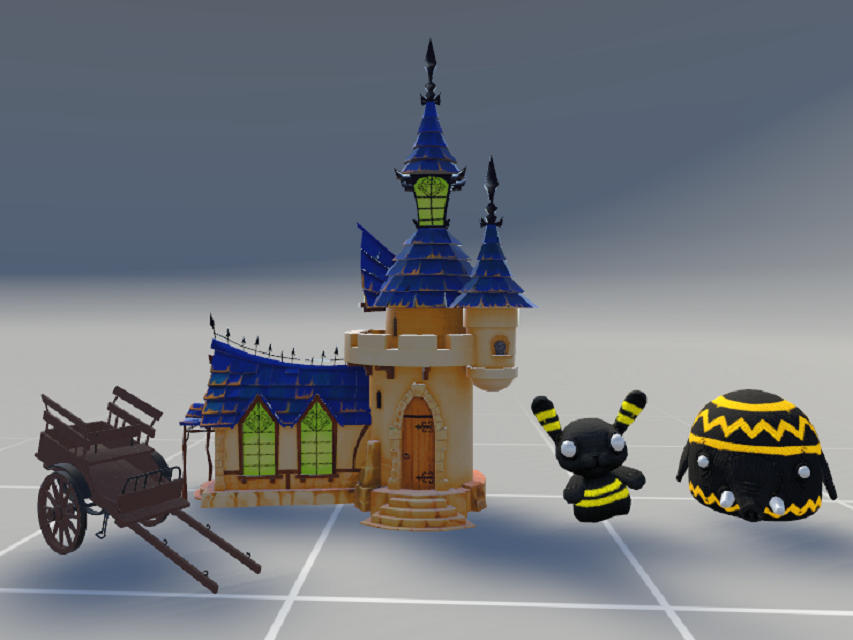}\label{fig:designs:c}}
  \subfigure[iconic design with negative emotion]{\includegraphics[width=0.19\textwidth]{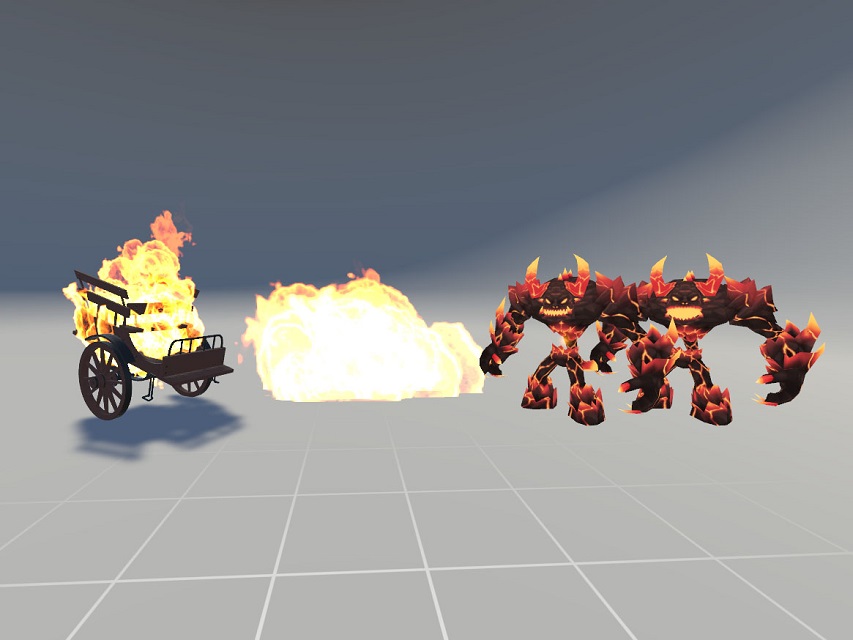}\label{fig:designs:d}}
  \subfigure[iconic design with neutral emotion]{\includegraphics[width=0.19\textwidth]{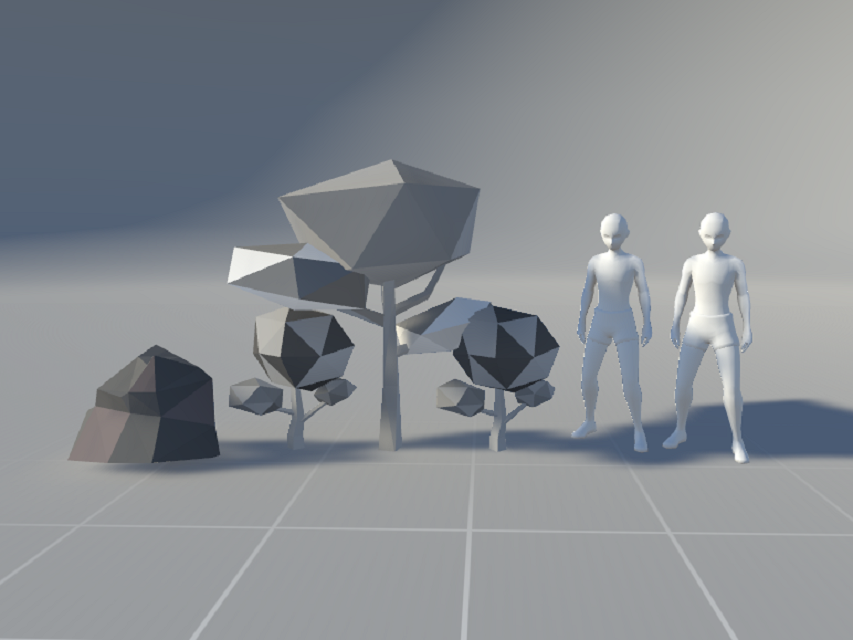}\label{fig:designs:e}}
  \caption{Five designs in different representational fidelities}
  \label{fig:designs}
\end{figure*}

\zhenyi{
As mentioned above, vividness and interactivity are the two important factors to presence. Considered that the environmental elements we plan to design do not belong to the VR experience originally, instead of stimulating users to revise them, we prefer prohibiting them from interacting with the elements. Hence we only keep vividness as one dimension of the design. We base our design ideas from aesthetic theories: imitationalism, formalism and expressionism~\cite{ragans2005art}, which share similar categories with the levels of representational fidelity: indexical, iconic and symbolic~\cite{kim2010design,pousman2006taxonomy}. As a result we generate our designs based on these three levels:
}

\textit{Imitationalism} and \textit{indexical} (\textbf{IDX}) refers to a direct high-fidelity representation of the target so that the viewer can immediately recognize the object. 

\textit{Formalism} and \textit{symbolic} (\textbf{SYB}) represent an abstract approach to presence. The target and its representation have a basic idea in common, but there is little clarity on what exactly they are. 

\textit{Expressionism} seeks to communicate a particular emotion when an object is perceived, while \textit{iconic} refers to a metaphor-based mapping between the target and its representation. 

\zhenyi{Apart from vividness, we define \textit{emotion} as another dimension for the fidelity. Instead of combining them together, we split only \textit{Expressionism} and \textit{iconic} representation into three emotional categories: positive (\textbf{POS}), negative (\textbf{NEG}) and neutral (\textbf{NEU}). For IDX design, such arrangement is more realistic and there is little room to develop different emotions. And for SYB design, ithe design is highly abstract and has too few details to exhibit emotions}.

\zhenyi{
The resulting designs following the rationale are shown in Figure~\ref{fig:designs}. 
Since furnitures such as table (a static object) and chair (a passively movable object) and bystanders (people proactively moving around) are representatives of different types of common environmental elements in VR experiences, 
we focused on designing the virtual presence of these three entities.
For the IDX style, we have a set of photo-realistic 3D models such as a realistic human being, a wooden table, and a wooden chair (Figure~\ref{fig:designs:a}). 
And Figure~\ref{fig:designs:b} shows a high-level abstract design in the form of a wire-frame box with a warning panel for the SYB style.
Then Figure ~\ref{fig:designs:c} displays a colorful, cute and vivid set of models including a castle, cart and animals for the POS style.
Figure~\ref{fig:designs:d} shows a dangerous-looking set of representations including fire and a demon for the NEG style.
And a simple colorless design with a low polygon counts for the NEU style (Figure~\ref{fig:designs:e}).
}



Since these representations are not strictly part of the game world, they do not interact or collide with the virtual objects in the game. For example, physics-controlled objects in the game world fall through these representations instead of landing on them or bouncing off them.

\xm{
To assess the efficacy of the five styles of design, we define the following the dependent variables:
\textbf{Situational awareness} could be measured at three levels, i.e., perception of the elements, comprehension of the current situation, and projection of future status. To be more specific, we captured perception by awareness attraction and attention retention on each element. Comprehension relates to how well users perceive the physicality, risk, and the need to avoid. Projection means the perceived ability to project an element's status (dangerous or not) in the future. We evaluated SA through 5-pt Likert scale rating in the questionnaires and post-study interview (as shown in Table~\ref{tab:questions} row 3 to 9)

\textbf{Perceived breaks in presence (BIPs)} is the subjective feeling of breaking feeling from immersive experience. We evaluated perceived BIPs in questionnaires on a 5-pt Likert scale. Additionally, we counted the number of different incidents of BIPs through observation, such as collision of a physical object, interruption of virtual task, and conversation with the bystander, to get a sense of the participants' actual experiences (as shown in Table~\ref{tab:questions} row 10).

\begin{table*}
  \centering
  \begin{tabular}{l l l }
    \textit{Measure} & \textit{Sub-measure} & \textit{Questionnaire Item}  \\
    \midrule
    \multirow{ 2}{*}{Manipulation Check} & Vividness & Q1: to what extent you think this design is vivid \\
     & Emotional Valence & Q2: to what extent you think the emotion of this design is positive\\
    \midrule\multirow{ 2}{*}{SA: perception} & Awareness Attraction & Q3: to what extent you were aware of the virtual representations\\
     & Awareness Retention & Q4: to what extent the virtual representations held your awareness\\
    \midrule\multirow{ 3}{*}{SA: comprehension} & Physicality & Q5: to what extent you think the representations have physical existence\\
     & Perceived Danger & Q6: to what extent you felt the virtual representations is dangerous\\
     & Avoidance & Q7: to what extent you intended to avoid the virtual representations\\
     & Interpretation & Q8$\star$: Why do you think the element shows up here\\
     \midrule\multirow{ 2}{*}{SA: projection}& & Q9: to what extent you think it will become a threat in future\\
     & & Q10$\star$: what do you expect to happen for the virtual representations \\
     \midrule
    Perceived BIP & & Q11: to what extent you felt breaks from the immersive experience\\
    \midrule
    Perceived Engagement & & Q12: to what extent you felt engaged during experience\\
    \bottomrule
  \end{tabular}
  \caption{Measurements and corresponding questionnaire items. Note that only Q8$\star$ and Q10$\star$ are open-ended questions. All the others are 5-pt Likert scale ratings (5 being the highest extent). }~\label{tab:questions}
\end{table*}

\textbf{Perceived engagement} refers to how involved users feel in the experience. We evaluated it through a questionnaire item on a 5-pt Likert scale \zhenyi{(as shown in Table~\ref{tab:questions} row 11)}.

In addition, we added manipulation check questions (5-pt Likert scale in Table~\ref{tab:questions} row 1 and 2) to verify whether the participants' perception of each design's vividness and emotion matches our design intention. 

These are our main dependent variables of interest, allowing us to gain a comprehensive understanding of how participants experience VR when the existence of physical elements around around is manifested in the virtual world.
}

\begin{figure}[ht]
  \centering
  \includegraphics[width=0.45\textwidth]{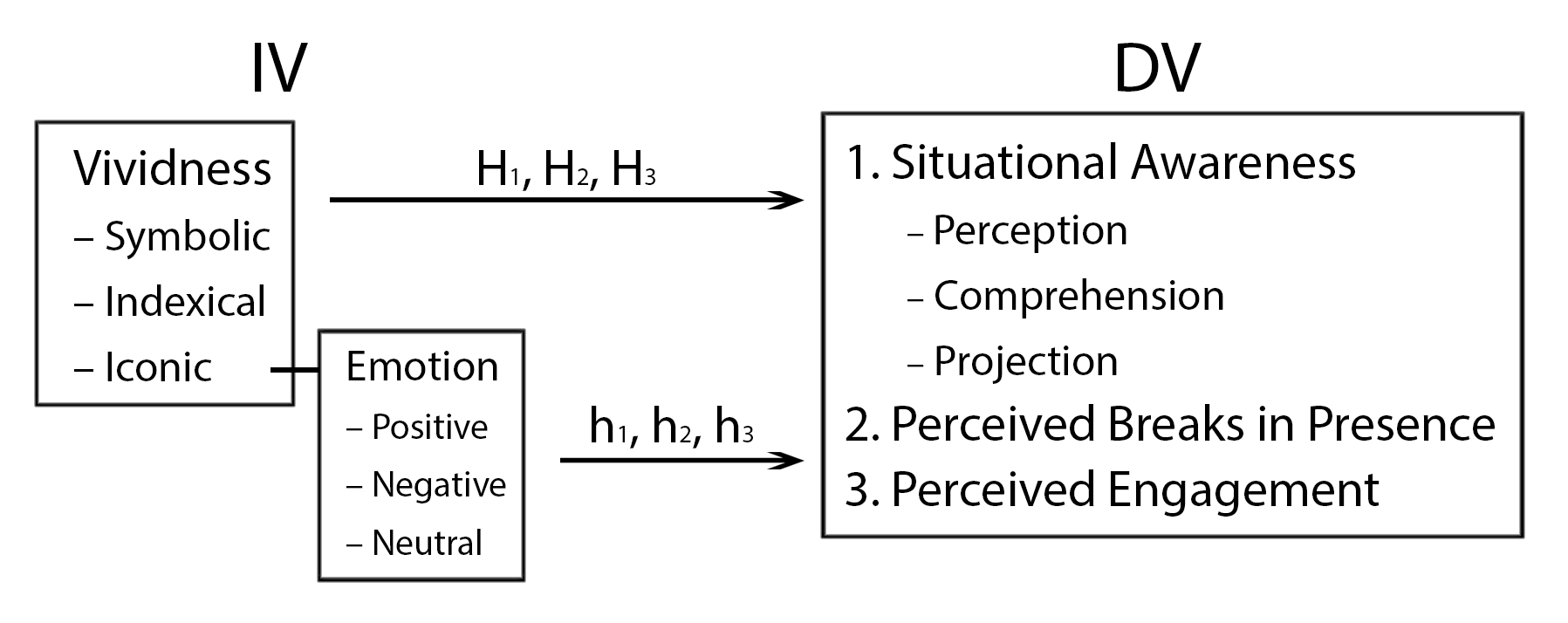}
  \caption{Research model and hypotheses}
  \label{fig:hypothesis}
\end{figure}

\xm{
\subsection{Research Model and Hypotheses}
Figure~\ref{fig:hypothesis} shows our proposed research model. More specifically, as vividness denotes the ability to introduce a sense a presence~\cite{steuer1992defining}, we hypothesize that:

\begin{itemize}
\item \textit{H1}. Fidelity with higher vividness (IDX) leads to significantly more SA than that with lower vividness (SYB), i.e., easier to gain (\textit{H1.1a}) and retain (\textit{H1.1b}) awareness of \zhenyi{representations},
better comprehension of the (\textit{H1.2a}) physicality, (\textit{H1.2b}) potential risk, and (\textit{H1.2c}) need to avoid of an element,
and (\textit{H1.3}) better projection of future status. \zhenyi{Among all of the above, iconic designs with median vividness are in the middle.}
\item \textit{H2}. Fidelity with higher vividness (IDX) leads to significantly fewer perceived BIPs than that with lower vividness (SYB). \zhenyi{In the middle there lies the Iconic designs with median vividness.}
\item \textit{H3}. Fidelity with higher vividness (IDX) is perceived significantly more engagement than that with lower vividness (SYB). \zhenyi{Similarly Iconic designs with median vividness are in the middle.}
\end{itemize}
}




\xm{Literature on the potential advantages of uncomfortable interaction (including negative emotion)~\cite{benford2013uncomfortable} suggests that discomfort in interaction design can help produce a more enlightening and focused experience. In spite of this, generally, negative designs tend to intimidate users, while positive affect may promote interactivity ~\cite{beale2008affect}. Hence we hypothesize that:
\begin{itemize}
\item \textit{h1}. Fidelity with negative emotion (NEG) will lead to more SA than positive (POS) and neutral (NEU) emotion, 
i.e., easier to gain (\textit{h1.1a}) and retain (\textit{h1.1b}) awareness of \zhenyi{representations},
better comprehension of the (\textit{h1.2a}) physicality, (\textit{h1.2b}) potential risk, and (\textit{h1.2c}) need to avoid of an element,
and (\textit{h1.3}) better projection of future status.
\item \textit{h2}. Fidelity with positive emotion (POS) will lead to less perceived BIPs than fidelities with the other emotions.
\item \textit{h3}. Fidelity with negative emotion (NEG) will lead to more perceived engagement than that with the other emotions. 

\end{itemize}
}


\zhenyi{
\subsection{Implementation of Prototype: CandyDream}
}
We aimed at creating a VR experience that:

\begin{enumerate}
\item{Allows and encourages users to physically move around in the space, and  simulates scenarios where users need to be aware of their physical surroundings during VR tasks.
}
\item{Provides a specific task with a clear goal for users to achieve, and prevents users from getting into a confusion situation and thus making unexpected decisions. 
} 
\item{Includes a learning session to ensure that users have mastered the task before proceeding to the main study. This is to eliminate any possible ordering effects, as the task is consistent across the five sessions. 
}
\item{Keeps the VR scene simple and controllable for evaluation.}
\end{enumerate}

\begin{figure}[h]
\centering
\subfigure[Hardware setup: 1) Vive trackers. 2) and 3) real physical furniture. 4) a bystander not using the VR system. 5) the user wearing a VR headset. 6) video camera. 7) Vive Lighthouse tracking system. ]{\includegraphics[width=0.2\textwidth]{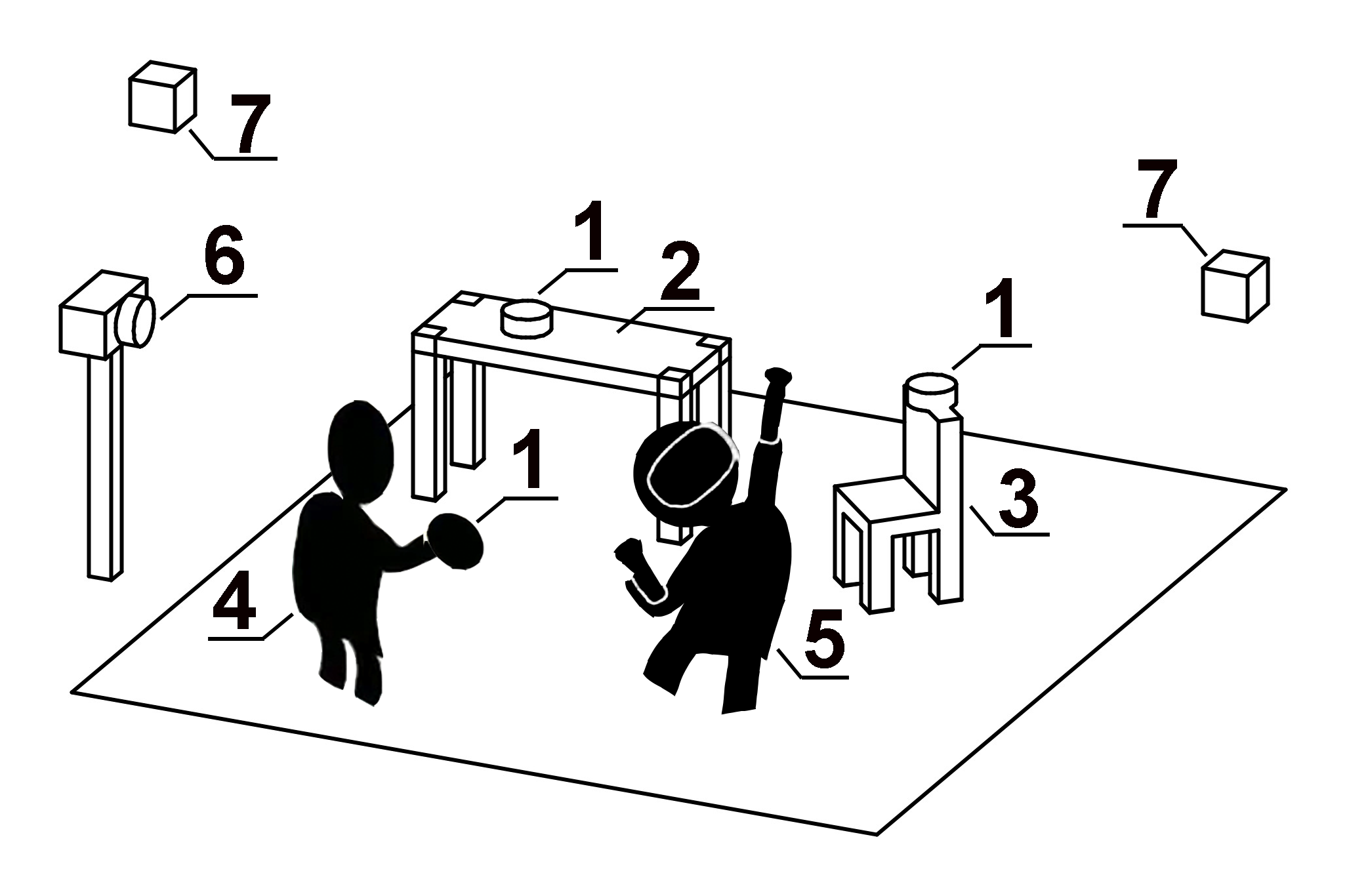}\label{fig:Hardware}}\quad
  \subfigure[Software setup: the user is highlighted \gabe{in} yellow, the virtual scoreboard and bucket for collection \gabe{are at the left side of the figure}. \gabe{Objects 1, 2, and 4 correspond to the same numbered objects in the hardware setup.}]{\includegraphics[width=0.2\textwidth]{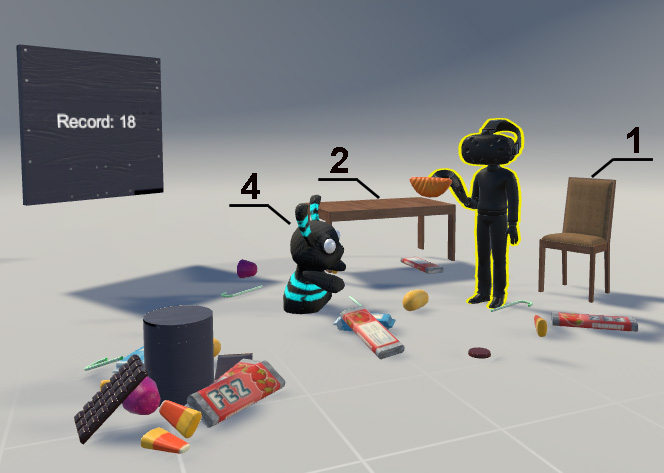}\label{fig:Software}}\\
  \caption{Hardware and software setup for CandyDream}
\end{figure}

Based on these criteria, We designed a VR demo, CandyDream, in Unity. 
The system setup is shown in Figure~\ref{fig:Hardware}.

Our physical set-up was an HTC Vive system in a 4.0m $\times$ 5.0m 
empty physical space. To \gabe{ease the user's learning curve}, we designed simple controls: performing the ``grab'' and ``throw'' actions in CandyDream requires only pressing and releasing the trigger button on one controller.

Users can throw a candy by mimicking a throwing motion with the controller, releasing the trigger button when they wish to release the candy. 
We used three extra Vive trackers to track the physical elements used in the study, including a 1.5m $\times$ 0.6m $\times$ 0.9m table as a static physical object, a 0.5m $\times$ 0.5m $\times$ 1.0m rollable chair as a movable physical object, and a bystander not wearing a VR headset. For safety reasons, we scaled all the virtual objects to be slightly bigger than their physical counterparts and added foam to all edges.
We used a powerful desktop computer (Intel(R) Core(TM) i7-4790K CPU @4.00GHz, 16GB RAM, SSD 840 PRO Series, GTX TITAN Black, Windows 10) to render the virtual world, thereby guaranteeing a smooth experience.

 \section{Evaluating Situational Awareness}
 \subsection{Apparatus and Participants}
We conducted a within-subject controlled experiment to study users' SA and potential BIPs under different designs. 

We recruited 25 participants (P1-P25) via email and word-of-mouth, including undergraduate and graduate students in either computer science or art and design from a local university, software engineers, and financial workers. The participants (44\% female) are between the ages of 18 and 32 (M=24.72, SD=3.13), and come from various countries, including the United States of America, China, Africa, and Saudi Arabia. According to the answers to the pre-screening questionnaire, 68\% have tried VR and 56\% have experienced HTC Vive. 

\subsection{Setup and Tasks}

One researcher serves as the investigator, in charge of instructing the study and taking notes. Another researcher plays the role of a normal "bystander" entering the interaction space. The bystander is instructed not to initiate contact with a user, and to avoid coming into contact or acting aggressively towards the user. 
We set up a video camera (as shown in Figure~\ref{fig:Hardware}) and a computer next to the experience space to record all the real-world events that happen among the participants, the physical objects (table and chair), and the bystander. We simultaneously record the screen of the Vive system.

Trail session and main session share the same game logic. At the beginning of the session, users are placed in a virtual world with different decorations depending on the session. During the session, virtual candies drop from the sky every 2 seconds and land randomly onto the virtual floor. 
To score in the task, users need to grab candies either in mid-air or on the ground with a handheld controller and throw the grabbed candies into a virtual bucket that is out of reach. The bucket stays at the same location (corresponding to the virtual scoreboard) across the five sessions. 
Each successfully collected candy counts as one point. 
The user's goal for trail is to earn 10 points before the time limit (two minutes) is up. 
The current score is displayed on a virtual scoreboard 
as shown in Figure~\ref{fig:Software}.

\subsection{Procedure}
After obtaining consent from the participants, we first introduce the game logic and interactions involved. 
Then we let the participants play a simple 2-minute trial of CandyDream in VR, in order to familiarize them with the task and environment without introducing our interventions. 
After a short break, participants proceed to the main task, \zhenyi{completing a 10-minute experience including} five 2-minute game sessions, each corresponding to one of the five designs. We counterbalance the order of the designs for each participant using Latin Square.
After each session, participants take a short break and complete a questionnaire about their experience with the associated design.
Note that during the breaks, participants do not get to see the objects or the bystander, so as to avoid leaking our intention to the participants (these elements only entered the room after participants put on the Vive Headset and started the game). The locations of the physical objects and bystander are also varied in each session to avoid participants making guesses based on knowledge gained from previous sessions.
Upon the completion of all sessions, participants will join a semi-structured exit interview, where they will provide feedback on their overall experiences.

 \section{Analysis and Results}



\begin{figure}[h]
\centering
\captionsetup{justification=centering}
	\includegraphics[width=0.4\textwidth]{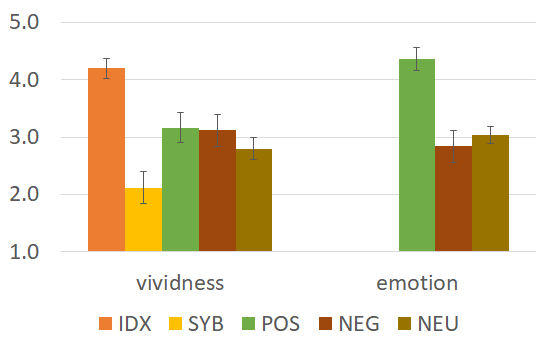}\label{fig:ux:mc}
  \caption{Means and standard errors of manipulation check for five designs on a 5-point Likert scale.}
  \label{fig:ux:1}
\end{figure}


\xm{
We conduct repeated measures MANOVA model on the quantitative results to test our hypotheses. We test the homogeneity of variance for each measure, and apply Greenhouse-Geisser correction when it is violated. We perform \zhenyi{Bonferroni}'s test for post-hoc analysis.
}

\subsection{Manipulation Check}

We check whether the designs are as expected based on \zhenyi{two dimensions} of representational fidelity (shown in Table~\ref{tab:questions} and Figure~\ref{fig:ux:1}).
First, analysis on our manipulation checks shows that \zhenyi{in dimension \textit{vividness}}, IDX is indeed perceived as \zhenyi{containing richer detail} (M=4.2, SD=0.865, F(4,96)=11.4, p<0.01) than the others. And SYB is the most abstract \zhenyi{and least detailed} (M=2.12, SD=1.43) of all. 
The results of other three designs in iconic category are in between as expected.

Then we gather the valence of \textit{emotion} among three \zhenyi{iconic} designs (shown in Table~\ref{tab:questions} and Figure~\ref{fig:ux:1}). 
POS is the most emotionally positive design (M=4.36, SD=1.04, F(2,48)=16.2, p<0.05).
And, NEG is perceived as the most negative design (M=2.84, SD=1.37). For example, one participant mentioned
\textit{``The fire element used in NEG is playful (rather than frightening as we anticipated), as it reminded me of previous game experiences''}(P12, female). Thus the reason NEG \zhenyi{does not receive \Karl{a very low score}} might be on account of this situation. 
Lastly, NEU (M=3.04, SD=0.73) is close to the middle of the scale. 
In summary, these five designs mostly met our expectation\Karl{s} in two dimensions.



\begin{table*}
  \centering
  \begin{tabular}{l l l r r r r r r}
    \textit{Dimension} & \multicolumn{2}{c}{\textit{DV}} & \textit{df} & \textit{MS} & \textit{F} & \textit{p} & \textit{${\eta}^2$} & \textit{Result} \\
    \midrule
    Vividness & SA: Perception & Awareness Attraction & 4 & 3.174 & 3.401 & 0.010 & 0.065 & \textbf{H1.1a} Accept\\
    & & Awareness Retention & 4 & 1.934 & 1.422 & 0.228 & 0.028 & \textbf{H1.1b} Reject\\
    & SA: Comprehension & Physicality & 4 & 8.126 & 5.740 & 0.020 & 0.105 & \textbf{H1.2a} Accept\\
    &  & Perceived Danger & 4 & 12.306 & 10.034 & 0.003 & 0.170 & \textbf{H1.2b} Reject\\
    &  & Avoidances & 4 & 1.400 & 0.879 & 0.478 & 0.018 & \textbf{H1.2c} Reject\\
    & \multicolumn{2}{l}{SA: Projection} & 4 & 11.590 & 12.875 & 0.001 & 0.208 & \textbf{H1.3} Partially Accept\\
    & \multicolumn{2}{l}{Perceived BIPs} & 4 & 2.610 & 2.926 & 0.025 & 0.109 & \textbf{H2} Reject\\
    & \multicolumn{2}{l}{Perceived Engagement} & 4 & 2.080 & 5.859 & 0.002 & 0.196 & \textbf{H3} Reject\\
    \midrule
    Emotion & SA: Perception & Awareness Attraction & 2 & 1.560&	1.975	&0.142	&0.026&	\textbf{h1.1a}  Reject\\
    & & Awareness Retention & 2 & 6.418&	5.167&	0.007&	0.065&	\textbf{h1.1b} Reject \\
    & SA: Comprehension & Physicality & 2	&0.871&	0.678&	0.509&	0.009&	\textbf{h1.2a} Reject\\
    &  & Perceived Danger & 2 &	38.298&	34.268&	$<$0.001&	0.317&	\textbf{h1.2b} Accept\\
    &  & Avoidances & 2	& 2.618 & 2.010 &	0.138	& 0.026 &	\textbf{h1.2c} Reject\\
    & \multicolumn{2}{l}{SA: Projection} & 2 &	28.093	& 33.763	& $<$0.001 &	0.313&	\textbf{h1.3} Accept\\
    & \multicolumn{2}{l}{Perceived BIPs} & 2&	1.013&	0.986&	0.380&	0.039&	\textbf{h2} Reject\\
    & \multicolumn{2}{l}{Perceived Engagement} & 2	&2.560	&7.580	&0.001&	0.240	&\textbf{h3} Accept\\
    \bottomrule
  \end{tabular}
  \caption{Results of hypotheses testing.}~\label{tab:hypoRes}
\end{table*}

\subsection{Situational Awareness}

Here we evaluate virtual representations of five designs from the perspective of SA. For dimension \textit{vividness}, we evaluated all five designs covering three levels of richness. For dimension \textit{emotion}, we evaluated POS, NEG and NEU to see how emotion of representation affect\Karl{s} SA.

\subsubsection{Level 1: Perception}
\begin{figure}[h]
  \centering
  \includegraphics[width=0.4\textwidth]{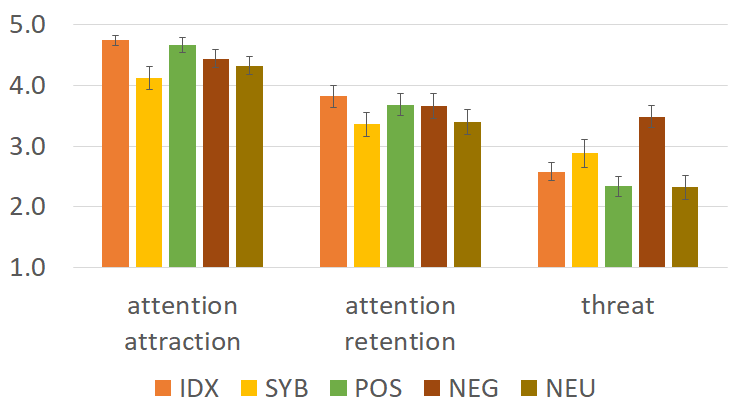}
  \caption{Means and standard errors of five designs of SA for perception and projection (threat) on a 5-point Likert scale.}
  \label{fig:sa:1}
\end{figure}

We asked the participants about attention attraction (Q3 in Table~\ref{tab:questions}).
In dimension \textit{vividness},
statistical results reveal significant difference\Karl{s} in terms of each representation's ability to draw users' awareness (see Figure~\ref{fig:sa:1} and Table~\ref{tab:hypoRes} row 1, \textbf{H1.1a} Accept).
According to the post-hoc pairwise comparison, IDX designs (M=4.74, SD=0.4) arouse significantly higher awareness than SYB designs (M=4.12, SD=0.95). Meanwhile the performance of POS, NEG and NEU are in between.
Thus the result indicates that the more vivid the design is, the more the design attracts the user. 
In dimension \textit{emotion}, there is no significant difference in this dimension (see Table~\ref{tab:hypoRes} row 9, \textbf{h1.1a} Reject).
And NEG designs (M=4.6, SD=0.32) attracted less attention than POS designs (M=4.68, SD=0.30) which rejects that emotion has strong impact to perception grasp. For example, one reminded that
\textit{``The bunny is really cute and large, and the demon is large as well, that I can't help but pay attention to them''(P18, female.).}
Based on the feedback and our observation, one possible explanation is that POS and NEG designs share similar richness in common, which is the key to attention attraction.

Then we asked the participants the question about attention retention (Q4 in Table~\ref{tab:questions}).
In dimension \textit{vividness}, statistical results reveal no significant difference in terms of each representation's ability to maintain users' awareness (see Table~\ref{tab:hypoRes} row 2, H1.1b Reject). 
One possible reason is that \zhenyi{participants lost their interest of exploring after they found these virtual representations were not interactive and helpful to task, such as \textit{``I did not pay too much attention after the very beginning because they did not affect my candy task. However, the elephant (bystander metaphor in POS) looked really adorable so that I came close to it sometimes''(P11, female).} That also indicated the richness of representation had poorer influence on attention retention than emotion did.}
In dimension \textit{emotion}, the result showed the significant difference (Table~\ref{tab:hypoRes} row 10, \textbf{h1.1b} Reject).
According to the post-hoc pairwise comparison, NEG (M=4.01, SD=0.75) keep more attention than POS (M=3.88, SD=0.70 , not significantly) and NEU (M=3.45, SD=0.82 ,p$<$0.05). Considering that NEU design hold less attention than both POS and NEG designs, one possible explanation is that attention is hold more by stronger emotion than by more negative emotion. 

During the experiments, we observed that six participants explicitly showed their fear towards the demon and stayed \Karl{alert the whole time to avoid it.}
\textit{``I just cannot move my eyes away from it. I need to know where it is in case it suddenly attacks me.'' (P5, male)}.
Meanwhile P4 (male) was fond of the metaphor, \textit{``It looks so funny especially when walking.''}
That also showed NEG has strong capability of holding attention.
In summary, vividness has positive influence on attention attraction, and emotion has weak impact on attention hold.

\begin{figure}[h]
  \centering
  \includegraphics[width=0.4\textwidth]{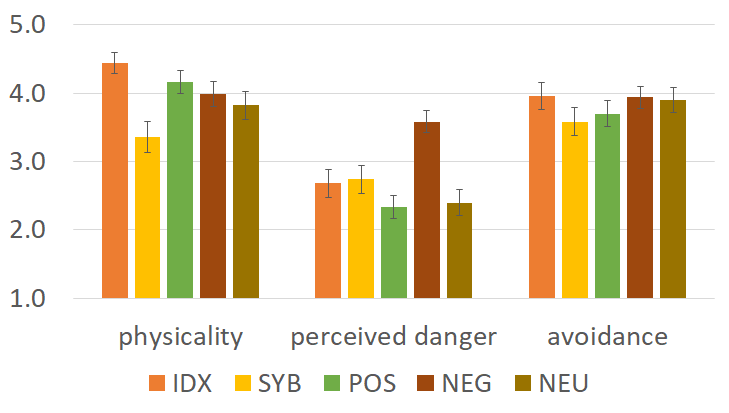}
  \caption{Means and standard errors of five designs of SA for comprehension on a 5-point Likert scale.}
  \label{fig:sa:2}
\end{figure}

\subsubsection{Level 2: Comprehension}
We asked the participants to rate the physicality (Q5 in Table~\ref{tab:questions}). The score will be high if the participants understand the existence of real physical elements during their VR experience.  
In dimension \textit{vividness}, analysis shows significant difference\Karl{s} in users' perceivability 
(see Table~\ref{tab:hypoRes}, \textbf{H1.2a} Accept).
According to post-hoc pairwise comparison, IDX received significantly more perceivability of physicality (M=4.44, SD=0.75) than SYB (M=3.36, SD=1.14, p$<$0.001). Also the iconic designs (POS, NEG and NEU) have score\Karl{s} in between (see Figure~\ref{fig:sa:2}).
The feedback \textit{``The moving wire-frame looks really weird and I totally have no idea what it is. (P1, male)''} also made the point that less richness of representation leads to less comprehension of their physicality.
\Karl{In dimension \textit{emotion}, there were no significant differences}
(see Table~\ref{tab:hypoRes}, \textbf{h1.2a} Reject and Figure~\ref{fig:sa:2}).
One possible reason is that POS and NEG designs are both far away from the real world so that participants treat them in similar ways no matter which emotion is triggered.
We can safely conclude that emotion has almost no help to participants' comprehension about the elements' physicality. \Karl{Emotion has almost no effect on participants' comprehension of the elements' physicality.}


We assessed the perceived risk through questionnaire (Q6 and Q7 in Table~\ref{tab:questions}).
For Q6, 5 means not dangerous at all. Elements under different designs evoked significantly different perception of risk 
(Table~\ref{tab:hypoRes}, \textbf{H1.2b} Reject)
in dimension \textit{vividness}.
Based on post-hoc pairwise comparison, IDX evokes less danger (not significantly) than SYB designs.
However, IDX does not receive less feeling of dangerous than all iconic (POS, NEG and NEU) designs (see Figure~\ref{fig:sa:2}).
Hence, the richness of the representations has very few impact here.
In dimension \textit{emotion}, there are still significant differences shown up (see Table~\ref{tab:hypoRes}, \textbf{h1.2b} Accept).
NEG designs (M=3.587, SD=0.43) received (significantly) higher danger perception than POS designs (M=2.41, SD=0.45) and NEU designs (M=2.29, SD=0.45) (see Figure~\ref{fig:sa:2}).
It is obvious that participants have more risky understanding about NEG design.

For Q7, 5 means no need to avoid. The results shows no significance among all elements 
(see Table~\ref{tab:hypoRes}, \textbf{H1.2c} Reject)
in \textit{vividness}, and likewise (see Table~\ref{tab:hypoRes}, \textbf{h1.2c} Reject) in \textit{emotion}. 
Combining what we heard from interview, users tend to adopt a more conservative strategy when they just start an immersive experience.\textit{ ``Unless the task has a special request for interaction, I prefer to not to touch the environment no matter it is part of the virtual game or real at the beginning.'' (P5, male)}.
Overall, the \textit{vividness} has some positive influence on comprehension. However, dimension \textit{emotion} has no influence here, no matter how strong the emotion is or how positive it is, except for perceiving risk.

We further collected the participants' interpretation of each element (Q8 in Table~\ref{tab:questions}).
It seems that 38.4\% of the participants did not have a clear idea of why these elements appeared in the game. Another 43.6\% shared the similar idea that these elements showed up in specific locations because there were real physical objects around, perhaps serving as a warning. 
The remaining 18\% had a different opinion. 
They thought that those content were all part of the game world.


\subsubsection{Level 3: Projection}

In this level, we assessed participants' projections in risk area and their prediction of the elements (Q9 and Q10 in Table~\ref{tab:questions}).

For Q9, 5 means that the element is very likely to become a threat. The analysis has significant difference (see Table~\ref{tab:hypoRes}, \textbf{H1.3} Partially Accept) in dimension \textit{vividness}. 
Based on post-hoc pairwise comparison, IDX designs (M=2.58, SD=0.98) projected less (not significantly) threat than SYB designs (M=2.88, SD=1.08). Also one iconic design (NEG) lead to higher threat than SYB (see Figure~\ref{fig:sa:1}).
In dimension \textit{emotion}, \Karl{a} significant difference remains (see Table~\ref{tab:hypoRes}, \textbf{h1.3} Accept). NEG design (M=3.57, SD=0.43) has \Karl{a} (significantly) higher score than POS and NEU (see Figure~\ref{fig:sa:1}).
Combining the feedback from participants, \textit{``The fire looked growing all the time and I felt like I will be engulfed very soon.'' (P10, male; P17, female).}
Thus we can tell that the cognition of threat is mostly from the dimension of \textit{emotion} and has weak connection with the \textit{vividness}.

Meanwhile there are some interesting finds for Q10. 52.8\% of the participants did not expect anything bad to happen. Another 27.2\% hoped to see some changes in the game scene based on game logic so that the game would be more vivid. And 20\% expected to have more interaction with the elements, either being attacked by them or playing with them proactively. In addition, for POS and NEG, 60\% of the participants expected to see changes brought by both designs. 
What's more, 20\% thought that the elements in NEG would attack them, which is much higher than other the results for other designs (M=3\%, SD=0.04). 
These findings suggest that participants anticipate that designs with more details and colors are more likely to change the game dynamics in the future, and they expect aggressive actions when seeing a NEG design.

\subsection{Perceived Breaks in Presence}

\begin{figure}[h]
  \centering
  \includegraphics[width=0.45\textwidth]{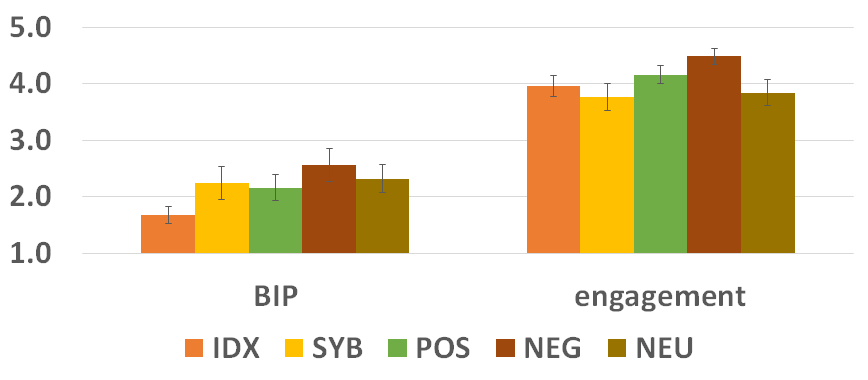}
  \caption{Means and standard errors of five designs of BIP and engagement on a 5-point Likert scale}
  \label{fig:bipengagement}
\end{figure}

We collected participants' feedback of BIPs based on Q11 in Table~\ref{tab:questions}. Feelings of BIPs evoke marginal difference among different designs (see Table~\ref{tab:hypoRes}, \textbf{H2} Reject) in dimension \textit{vividness}.
IDX has \Karl{does not have (significantly) lower} BIP (M=1.68, SD=0.75) than SYB (M=2.24, SD=1.48). Meanwhile, some iconic designs (NEU and NEG) have larger BIP than SYB do (see Figure~\ref{fig:bipengagement}). So \textit{vividness} has only weak influence on perceived BIPs.
In dimension \textit{emotion} (Table~\ref{tab:hypoRes}, \textbf{h2} Reject), there is no significant difference among these three designs.
One possible explanation might be the size of the representation, \textit{"I felt the bunny (from POS) and the demon (from NEG) both occupied a lot of space which reminded me their existence and stopped me from finishing my task"(P11, female)}.

Although we do not find any significant differences in terms of perceived BIPs in the proposed dimensions, we indeed observed several types of BIP incidents during the experiments.

\textit{Collision}.
We observed that 28.8\% of the participants knocked into the static objects during the experience, while 12\% collided with the bystander. Among all the participants, \Karl{the} table and chair in POS received the largest number of collision\Karl{s}. P11 (female), P22 (female) and P24 (male) share similar thoughts, \textit{``The models looked so fantastic that I do not think they were real.''}
And P2 (female), \textit{``I only realized the metaphor represented a real object after the collision.''}

\textit{Interruption}.
During the tasks, some participants suddenly realized that someone was nearby. If they \Karl{did} not \Karl{perceive} that the bystander metaphor around was that of a real bystander, the experience was strongly broken.

\textit{Rest}.
We observed an interesting phenomenon that three participants (P4, P8 and P24) sat on the chair when they figured out the physicality of the chair, which slightly showed that users' experience\Karl{s} were broken to some extent.
\textit{``Because I feel there is a real-world correspondence with the VR object, so I think I can sit on the chair.''(P8, male)}.

In the post-study interviews, we ask the participants for means to avoid these BIP incidents. One theme that emerges (mentioned by 72\% of the participants) is that \textit{``good VR content should adopt a consistent art style, even for VR presence designs of physical objects, to minimize the feeling of breaks.''(P7, male)}. This suggests that designs more consistent with the look of the original VE might help reducing the BIPs.

\subsection{Perceived Engagement}
We asked participants to rate their engagement from 1 to 5 among five designs (Q12 in Table~\ref{tab:questions}).
The results show significant difference\Karl{s} in \textit{vividness} dimension (Table~\ref{tab:hypoRes}, \textbf{H3} Reject).
The perceived engagement of IDX (M=3.96, SD=94) is slightly (not significantly) higher than SYB (M=3.76, SD=1.17) as hypothesized.
However we found NEG received the highest score (M=4.48, SD=0.72, compared to SYB and NEU; p<0.05). Thus we think vividness has very few influence on engagement perceived.
To dimension \textit{emotion}, there shows significant difference (Table~\ref{tab:hypoRes}, \textbf{h3} Accept). 
Based on post-hoc pairwise comparison, NEG received (significantly) higher engagement than NEU (M=3.84, SD=0.68) and (marginally) higher engagement than  POS (M=4.16, SD=0.48).
Combining the feedback of NEG design, 
\textit{``I feel nervous in NEG and I was highly focused and engaged into the experience then.'' (P5, male).
Meanwhile, another user (P17, female) with rich game experience, ``NEG is really close to what I usually play, and it felt familiar.''}
So that we can tell \textit{emotion} does affect the engagement, and the \textit{vividness} has weak influence here. 
 
 \section{Discussion}
\mxj{
\subsection{Comparison across Table, Chair, and Bystander}
}
\zhenyi{
During the analysis we found some specific elements have different behavior from their design categories.
Although the whole NEG designs do not have significant difference when comparing to POS, bystander metaphor in NEG (M=4.72, SD=0.324) holds the most attention and the score is significantly larger than that of chair metaphor (M=3.52, SD=1.5) in POS. Hence, combining the feedback \textit{"The demon is moving all the time which I can't move my eyes away from it." (P11, female)}, we can safely conclude that the moveability of the element have effects to some specific conditions such as attraction attention. 
\xm{
\subsection{Effect of Having a Rich Representation}
Our results show that providing rich details in the design of a virtual presence makes it easier for participants to be and stay aware of the corresponding element. This is because fidelity vividness may affect humans' perception of presence in two aspects: realism and believability~\cite{casati2005subjective}. Realism ``intends to approximate the model, the real world, in a very accurate way'', while believability only provide details ``considered relevant or representative of the intention behind the model'' ~\cite{casati2005subjective}. On one hand, if a representation stresses cues of realism, users are more likely to map it to the physical world. \zhenyi{\textit{``Once I saw the table and chair (metaphor in POS), I immediately believe they are real ones. And that is why I sit on the chair during the rest of this session.''(P4, male)}.} On the other hand, if a representation affords believability, or the feeling that it goes with the virtual scene, users may experience fewer BIPs. 

Additionally, it is essential to balance these two aspects in the design, to avoid falling into the Uncanny Valley~\cite{brenton2005uncanny}. That is to say, the design should not create doubts about the virtual presence of the elements that users would find disturbing.
\zhenyi{\textit{``The woman I met in the fourth session (IDX) looks so unreal to me. It has too few detail as a human and I am still confused about what it exactly is.''(P12, female)}}



\subsection{Effect of Stimulating Emotions in Design}
Although our hypotheses on situational awareness and perceived breaks in presence along the dimension of \textit{emotion} were not fully accepted, we found that designs that induce emotion, regardless of the valence, performed significantly different from the NEU design along some of the measures. 
In other words, the ability of a design to stimulate user emotion may matter more than the actual direction of induced emotion in some cases.

We also notice that, although NEG could arouse immediate awareness of an environmental element, users may update their belief as they proceed in the task. ``\textit{I felt worried at the beginning. However, after a few attempts to verify their identities, I realize that there were physical objects behind the representations. Then I feel OK to interact with them.}'' (P4, male; P18, female). 

Despite that an emotion-inducing design may not lead to more severe BIPs than a neutral design, they may sidetrack users for a longer time.  In addition, many users express that they have more interests in interacting with the POS/NEG environmental factors than the core mission. If the goal is to keep users focused on the main task in the virtual world, it is better to keep the representation of environmental factors neutral. \zhenyi{\textit{``The first one I tried (NEU) has very few color. Sort of I felt no happiness or fear during the task and I chose to ignore them (the elements) for most of the time.''(P14, female)}.}

}
}

\subsection{Limitations of this work}

From a scenario perspective, we did not implement a game scene with rich details and interactions for the sake of simplicity. When VR content and virtual presence of physical surroundings are presented together, the situation becomes complicated and the results might shift.
In addition, we only designed five representational fidelities and evaluated them separately. However, the design can be a mix across different designs. For example, SYB can be added up to any iconic design. When people approach those objects, a simple mesh net should be shining, thus providing a hint of the existence of physical surroundings.
 
 \section{Conclusion and Future Work}
In this paper, we designed five representational fidelities (indexical IDX, symbolic SYB, and iconic with three emotions -- POSitive, NEUtral, and NEGative) from two dimensions for displaying ambient entities in VR based on the existing aesthetic theory. We evaluated the efficacy of these five designs in terms of raising users' SA, perceived BIPs and perceived engagement by a within-subject study with 25 participants.

\zhenyi{Results show that designs with higher vividness evoke more SA, designs that are more consistent with the virtual environment can mitigate the BIP issue, and emotion-evoking designs are more engaging.}

In the future, we will evaluate our design in a more complex VR environment with more diverse tasks to verify the scalability and generality of our findings.
\zhenyi{And we will also investigate how the intensity of emotion affects SA. }
In addition, we plan to conduct a study in the wild to see how participants may perceive and handle real risks, and evaluate whether our design can truly alleviate psychological discomfort in such settings.

\bibliographystyle{IEEEtran}
\bibliography{sample}


\end{document}